\documentclass[aps,prb,twocolumn,superscriptaddress,showpacs]{revtex4}
\usepackage{graphicx}

\def\be{\begin{equation}}
\def\ee{\end{equation}}
\def\ba{\begin{eqnarray}}
\def\ea{\end{eqnarray}}

\def\bc{\begin{center}}
\def\ec{\end{center}}

\begin{document}

\title{Observation of exchange Coulomb interactions in the quantum Hall state at $\nu=3$}

\author{A.~B.~Van'kov}
\affiliation{Institute of Solid State Physics, RAS, Chernogolovka,
142432 Russia}
\affiliation{Department of Physics, Columbia
University, New York, New York 10027, USA}

\author{T.~D.~Rhone}
\affiliation{Department of Physics, Columbia University, New York,
New York 10027, USA}

\author{A.~Pinczuk}
\affiliation{Department of Physics, Columbia University, New York,
New York 10027, USA} \affiliation{Department of Applied Physics and
Applied Mathematics, Columbia University, New York, New York 10027,
USA}

\author{I.~V.~Kukushkin}
\affiliation{Institute of Solid State Physics, RAS, Chernogolovka,
142432 Russia}

\author{Loren N. Pfeiffer}
\author{Ken W. West}
\affiliation{Bell Laboratories, Alcatel-Lucent, Murray Hill, New
Jersey 07974, USA\footnote{Present address: Department of Electrical
Engineering, Princeton University, Princeton, NJ 08544, USA}}

\date{\today}

\begin{abstract}
Coulomb exchange interactions of electrons in the $\nu=3$ quantum
Hall state are determined from two inter-Landau level spin-flip
excitations measured by resonant inelastic light scattering. The two
coupled collective excitations are linked to inter-Landau level
spin-flip transitions arising from the N=0 and N=1 Landau levels.
The strong repulsion between the two spin-flip modes in the
long-wave limit is clearly manifested in spectra displaying Coulomb
exchange contributions that are comparable to the exchange energy
for the quantum Hall state at $\nu=1$ . Theoretical calculations
within the Hartree-Fock approximation are in a good agreement with
measured energies of spin-flip collective excitations.

\end{abstract}


\maketitle

The exchange Coulomb interaction energy of electrons on Landau
levels (LL) plays key roles in quantum Hall systems, particularly at
odd values of the filling factor, $\nu=nhc/eB$ (where $n$ is the
areal density), when the 2D electron system evolves into a quantum
Hall ferromagnet. One way to probe the exchange interaction is by
measurements of energies of collective spin-flip excitations. The
simplest one is the spin-wave, in which Landau orbital quantization
does not change. At odd filling factors the spin wave energy in the
short wavelength limit is predicted to have a large exchange
contribution, resulting in an enhanced spin
gap~\cite{by81,kallin84}. However, the actual energy values measured
in activated transport experiments turned out to be significantly
below theoretical estimates. These discrepancies occur in both the
integer and fractional quantum Hall regimes. Possible reasons for
the discrepancies lie in impact of spin-textures (skyrmions)~\cite{
schmeller,dethlef,grosh}, and of weak residual disorder potential
~\cite{usher,dolgo,khrapai}.
\par
Experimental venues to access Coulomb exchange interactions also
emanate from determinations of collective excitation modes in the
spin degree of freedom. At odd integer values of filling factor
the long-wavelength spin-wave is a minimum energy collective
excitation. The long-wavelength spin-wave mode approaches the
unshifted Zeeman energy~\cite{by81,kallin84,Larmor} and carries
marginal information about the electron-electron interaction.

In contrast, inelastic light scattering methods enable the direct
determination of exchange Coulomb interactions from measurements of
spin-flip collective excitations across cyclotron gaps
~\cite{nu1pinczuk,nu1vankov,nu1temp}. In these spin-flip (SF)
excitations there is simultaneous change in Landau quantization and
in orientation of spin. The long wavelength SF excitations represent
probes that are nearly insensitive to perturbations on length scales
exceeding the characteristic size of the quasiparticle-quasihole
pair magnetoexciton that is of the order of the magnetic length
$l_o=(\hbar c/eB)^{1/2}$, where $B$ is the perpendicular component
of magnetic field.

At $\nu=1$ the electron-electron interaction affects the energy of
the long-wave cyclotron SF mode, which involves the change of the
Landau quantization number by +1. This mode is shifted upwards from
the cyclotron energy by about half the full exchange energy in the
large momentum (small wave length) spin wave. Studies of the
cyclotron SF mode at $\nu=1$ have shown that the Coulomb exchange
contributions to its energy scale as $\sqrt{B}$ and its value is
softened by the spread of the electron wave-function in the
direction normal to the 2D-plane. Theoretical predictions are in
good agreement with measured mode energies determined as function of
magnetic field and quantum well width ~\cite{nu1pinczuk,nu1vankov}.
\par
We report inelastic light scattering measurements of collective
inter-Landau level excitations in the quantum Hall state at
$\nu=3$. All collective excitations are identified in spectra of
inelastic light scattering and their energies are compared with
theoretical calculations. We identified two coupled cyclotron SF
modes arising from the N=0 and N=1 Landau levels and interpreted
the results in terms of Coulomb exchange interactions. We
determined that these coupled cyclotron SF modes at $\nu=3$ are
subject to large Coulomb exchange interactions that are comparable
to the exchange energy in the quantum Hall state at $\nu=1$.
\par
There is great current interest in the roles of the spin degree of
freedom in the remarkable quantum Hall phases that emerge in the N=1
Landau level ~\cite{csathy,dean,trevor}. The finding reported here
is that exchange Coulomb interactions in the N=1 Landau level are
comparable to those in the N=0 level. This comparatively simple
result suggest that the exotic collective states that emerge in the
partially populated N=1 level are linked to the differences in
correlation effects between the two levels.
\par
Figure\,1a shows the schematic representation of five lowest energy
collective excitations in case of filling factor $\nu$=3. They are
shown as magnetoexcitons consisting of an electron promoted from a
filled Landau level and bound to an effective hole left in the
``initial'' LL. This representation is exact in the limit of strong
magnetic field where the parameter $r_c=E_c/\hbar \omega_c$ is small
enough~\cite{by81,by83,kallin84}. Here $E_c$ is the characteristic
Coulomb energy scale and $\hbar \omega_c$ is the cyclotron energy.
The set of dispersion curves of the collective modes can be
described in the following way~\cite{kallin84}:
\begin{equation}
 E_{m, \delta S_z}(k)=m \hbar
\omega_c + g \mu_B B \delta S_z + \Delta E_{m, \delta S_z}(k),
\end{equation}
where $m$ is the change in the LL index, $g \mu_B B \delta S_z$ is
the bare Zeeman energy associated with spin-flip. The last term
$\Delta E_{m, \delta S_z}(k)$ is alone responsible for the
dispersion and comprises all contributions from the many-body
Coulomb interaction and exchange energies in the initial and the
excited states. In the present discussion we focus on the excitation
spectrum with $m=0$ and $m=1$.
\par
At $\nu=3$ the four inter-LL transitions with $m=1$ shown in
Fig.1a are not independent. They couple via the Coulomb
interaction to yield two pairs of excitations. For the two
inter-LL excitations with changes in the charge degree of freedom
with $\delta S_z=0$ we have the in-phase magnetoplasmon (MP) mode
and the antiphase plasmon (AP) mode. For the the two excitations
with changes in the spin degree of freedom with $\delta S_z=-1$
the two coupled modes are cyclotron spin-flip excitations SF1 and
SF2.
\par
In first-order perturbation theory the dispersion curves of the
coupled modes are expressed as follows:
\begin{eqnarray}
E_{1,2}(k) & = & \frac{{\cal E}_1(k)+{\cal E}_2(k)}{2} \pm
 {}\nonumber\\
 & \pm & \sqrt{\left( \frac{{\cal E}_1(k)-{\cal E}_2(k)}{2}
\right)^2 + \Delta_{12}(k)^2},
\end{eqnarray}
where ${\cal E}_{1,2}(k)$ are the energies of single transitions
either with or without spin-flip, $\Delta_{12}(k)$ -- is the term,
responsible for coupling. For MP and AP excitations this theory
yields a vanishing Coulomb term $\Delta E(k)$ in the
long-wavelength limit. Unlike MP, for which the Kohn's
theorem~\cite{Kohn} is valid, the experimental values of the
energy of AP mode are red-shifted relative to the cyclotron energy
at integer filling factors $\nu \geq 2$. The experimental results
were reported in
Refs.[\onlinecite{nu2pinczuk,nu2chg,DrozdovKulik}] and the
explanation was given in the framework of the second-order
perturbation theory~\cite{nu2dickm,DrozdovKulik}.
\par
We calculated the wave vector dispersions of SF1 and SF2 at
$\nu=3$ in terms of matrix elements $\tilde{V}^{(1)}_{\alpha \beta
\gamma \delta}(k)$ introduced in Ref.[\onlinecite{kallin84}]:
\begin{eqnarray}
 {\cal E}_1(k) & = & \hbar \omega_c + \left | g \mu_B B \right | +
\Sigma_{0\,\uparrow,1\,\downarrow}- \tilde{V}^{(1)}_{1001}(k)\\
{\cal E}_2(k) & = & \hbar \omega_c + \left | g \mu_B B \right | +
\Sigma_{1\,\uparrow,2\,\downarrow}-
\tilde{V}^{(1)}_{2112}(k)\nonumber\\
\Delta_{12}(k) & = & \tilde{V}^{(1)}_{1102}(k)\nonumber
\end{eqnarray}
where
$\Sigma_{0\,\uparrow,1\,\downarrow}=\tilde{V}^{(1)}_{0000}(0)+\tilde{V}^{(1)}_{0101}(0)-\tilde{V}^{(1)}_{1010}(0)$
and
$\Sigma_{1\,\uparrow,2\,\downarrow}=\tilde{V}^{(1)}_{1010}(0)+\tilde{V}^{(1)}_{1111}(0)-\tilde{V}^{(1)}_{2020}(0)$
are the differences of exchange-self energies in the excited and
ground states for two single spin-flip transitions between
adjacent LLs depicted on Fig.1a. The calculated dispersion curves
for all four inter-Landau level excitations at $B_{\perp}=$5.3\,T
are plotted on Fig.1b by solid lines. For comparison with the
experiment, performed on the 24nm quantum well, it was essential
to take into account the finite thickness of the 2D-electron
system. For this the Fourier component of the effective {\it e-e}
interaction potential $\vartheta(q)=2\pi e^2/\varepsilon q$ was
multiplied by the geometric form-factor $F(q)$ calculated via the
self-consistent solution of the Poisson's and Schr\"{o}dinger's
equations~\cite{lu93}.

\begin{figure}[htb!]
\includegraphics[width=.48\textwidth]{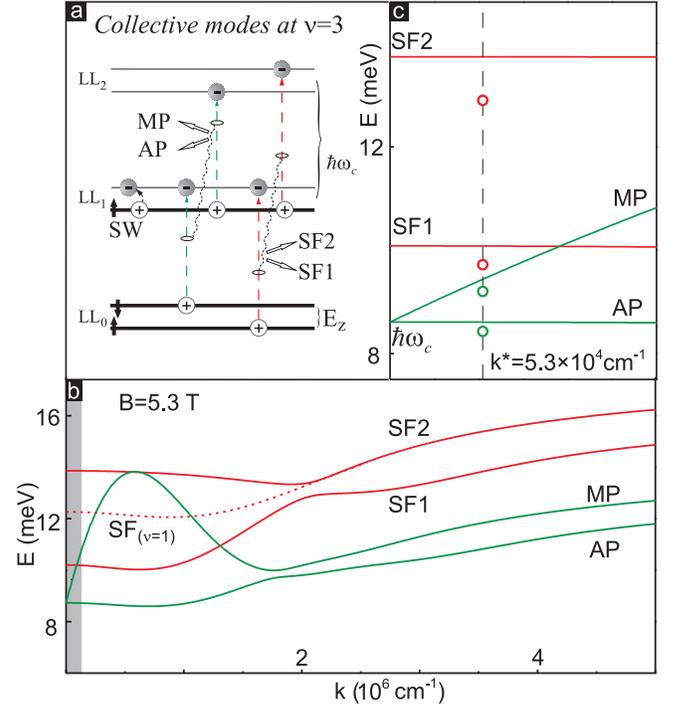}
\caption{\label{fig1} (a): Schematic representation of the
formation of collective modes at $\nu=3$ from single-electron
transitions. Spin-wave (SW) is described as a single spin-flip
transition within half-filled LL$_1$. MP and AP are formed as
inphase and antiphase combinations of two inter-LL transitions
with $\delta S_z=0$ (painted in green). Cyclotron spin-flip modes
SF1 and SF2 arise from analogous combinations of inter-LL
transitions with $\delta S_z=-1$ (painted in red). (b): Dispersion
curves of inter-LL excitations calculated at $B_{\perp}=5.3$\,T
within the first-order Hartree-Fock approximation are shown. Here
the finite thickness of the 2D electron system is taken into
account via the geometric form-factor. Dashed line represents the
dispersion of cyclotron spin-flip mode at $\nu=1$ and the same
magnetic field. (c): The zoomed-in image of the long-wavelength
region of Fig.1b, painted in light grey. The dashed vertical line
indicates the experimental in-plane momentum ${\rm k}^*=5.3 \times
10^4\,{\rm cm}^{-1} $. Open circles represent the experimental
data.}
\end{figure}

Both cyclotron spin-flip modes at $\nu=3$ are significantly
blue-shifted from the cyclotron energy and are nearly
dispersionless in the long-wavelength limit (Fig.1b,c).
Furthermore, they strongly repulse each other especially at small
momenta. As a result, the Coulomb energy of SF2 in the long
wavelength limit is even larger than that of analogous inter-LL
spin-flip mode in a fully spin-polarized quantum Hall state
$\nu=1$~\cite{nu1pinczuk,nu1vankov}. The Coulomb energy of the
long-wavelength mode SF2 is just 15\% smaller than that of spin
wave at $k \to \infty$ (shown on the inset to Fig.2) being the
exchange energy of electrons on the LL$_1$. On the contrary, the
energy of SF1 proves out to be pushed down. One of the intriguing
results of this calculation is that the highest energy spin-flip
excitation SF2 corresponds to the antiphased combination of two
single electron transitions and SF1 corresponds to the inphase
combination. In this aspect the situation is opposite to the case
of $\delta S_z=0$ modes MP (inphase) and AP (antiphase). As was
shown in case of AP~\cite{nu2chg,nu2dickm} the first-order
perturbation theory gives somewhat overestimated energy in the
long-wavelength limit. Although second-order corrections are
exactly computed only for AP at $k=0$, they are likely to be of
the same order also for SF1 and SF2.
\begin{figure}[htb!]
\includegraphics[width=.48\textwidth]{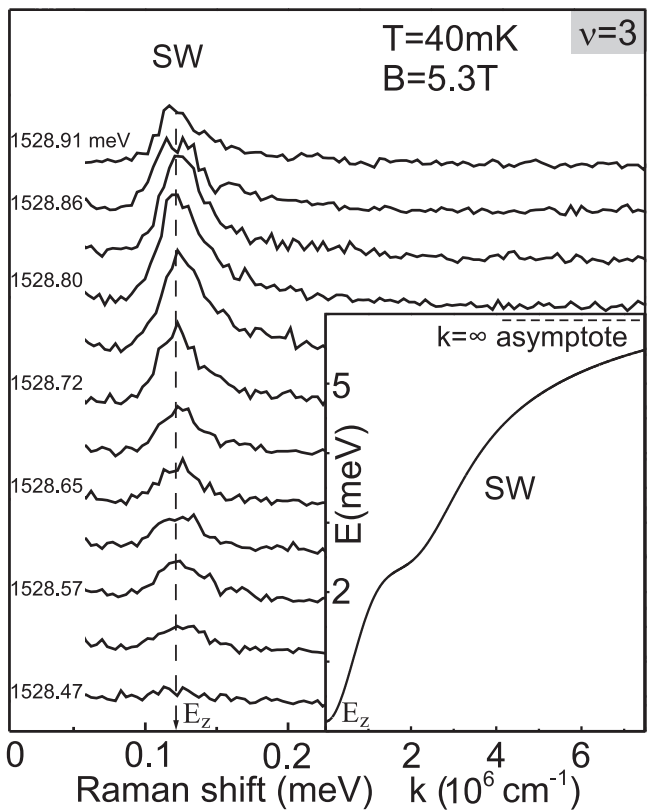}
\caption{\label{fig2} Inelastic light scattering spectra of
intra-LL SW mode at $\nu = 3$ and $B=5.3\ {\rm T}$ taken at
different laser photon energies (shown on the left). The inset
shows the SW dispersion curve calculated within the Hartree-Fock
approximation \cite{kallin84}, for a 24nm- wide quantum well. At
the experimental in-plane momentum the energy of SW is
indistinguishable from ${\rm E_Z}$, shown by a dashed arrow.}
\end{figure}

The inelastic light scattering measurements were performed on a
high quality ${\rm GaAs}/{\rm Al}_{0.3}{\rm Ga}_{0.7}{\rm As}$
single quantum well of width 24\,nm. The electron density is
$n=3.85 \times 10^{11}$\,cm$^{-2}$ and low temperature mobility
above $17 \times 10^6$\,cm$^2$/V$\cdot$\,sec. The sample was
mounted on the cold finger of a ${}^3{\rm He}/{}^4{\rm He}$
dilution refrigerator that is inserted in the cold bore of a
superconducting magnet. The refrigerator is equipped with windows
for optical access. Cold finger temperature was held mostly at
$T=40$\,mK or 1.7\,K for one part of the experiment. The
backscattering geometry was used at an angle $\theta=20^\circ$
with the normal of the sample surface. The perpendicular component
of the magnetic field is $B=B_T$\,cos$\theta$ and $B_T$ is the
total magnetic field. Resonant inelastic light scattering spectra
were obtained by tuning the incident photon energy of a
Ti:sapphire laser close to the fundamental optical gap of GaAs to
enhance the light scattering cross section. The power density was
kept below $10^{-4}$\,W/cm$^2$ for the measurements at
temperatures around 40\,mK. The in-plane momentum, transferred to
the excitations at the employed experimental geometry was about
$5.3 \times 10^4$\,cm$^{-1}$. The scattered signal was dispersed
by a triple grating spectrometer T-64000 working in additive and
subtractive modes and analyzed by a charge-coupled device camera.
The combined resolution of the system was about 0.02\,meV. In
order to distinguish between inelastic light scattering and
luminescence lines in spectra, the special test was employed --
when varying the incident photon energy, inelastic light
scattering lines traced the laser path, while luminescence lines
did not change their spectral position.
\par
The resonant enhancement of the intensities of light scattering
spectra of the long wavelength spin-wave (SW) mode at $\nu=3$ is
displayed in Fig.2. The SW is at the bare Zeeman energy with
$\left | g \right |=0.37$ and corresponds to the leftmost part of
the mode dispersion shown in the inset to Fig.2. Very small
changes in the laser photon energy (by $\sim0.5$\,meV)
dramatically affect the line intensity, indicating the importance
of resonant excitation in these experiments. The strong SW seen in
Fig.2 is consistent with the ferromagnetic character of the
quantum Hall state at $\nu=3$.
\par
Similar resonant excitation conditions prevail in the observation
of the inter-Landau level excitations reported below. To capture
light scattering spectra of inter-LL excitations, the incident
photon energies were chosen in such a way as to excite electrons
from the valence band to the second or third Landau levels in the
conduction band. Typical spectra are measured at two laser
positions (1538.2\,meV and 1550.0\,meV) shown on Fig.\,3.
\par
The magnetoplasmon and antiphased plasmon are seen shifted from
the cyclotron energy $\hbar \omega_c$=8.65\,meV (depicted by an
arrow on Fig.3). The blue shift of the MP results from the
2D-plasma energy at the non-zero in-plane momentum used in the
experiment. In fact, the MP is the only dispersive mode in the
range of experimentally accessible momenta (see Fig.\,1c). The
energy of AP is below the CR by 0.19\,meV. This energy shift is
somewhat smaller than that measured for 18\,nm quantum well in
Ref.[\onlinecite{DrozdovKulik}], which is a consequence of the
strong dependence of the effective Coulomb interaction strength on
the quantum well width. Developed in
Ref.[\onlinecite{DrozdovKulik}] theory gives $\Delta
E_{\rm{AP}}(0) \approx-0.25$\,meV for this magnetic field and
quantum well width.
\begin{figure}[htb!]
\includegraphics[width=.48\textwidth]{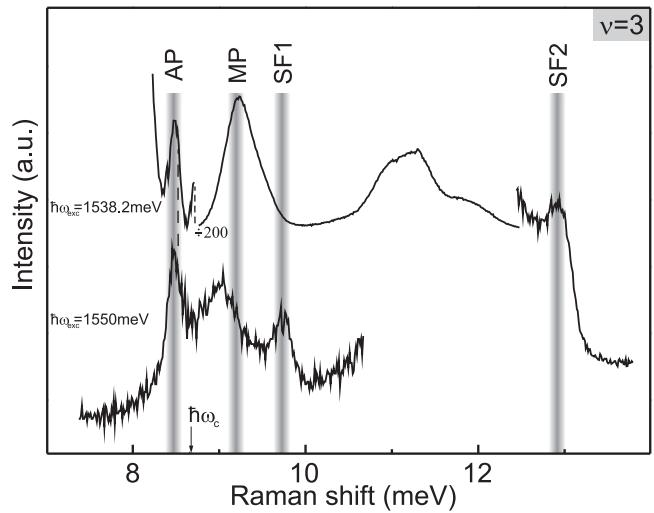}
\caption{\label{fig2} Inelastic light scattering spectra of
inter-LL excitations at $\nu = 3$ and $B_{\perp}=5.3\ {\rm T}$.
Upper spectrum is taken at the incident photon energy $\hbar
\omega_{exc}=1538.2\ {\rm meV}$, the lower one -- at 1550.0\,meV.
The position of $\hbar \omega_{c}$=8.65\,meV is shown by the
arrow. Grey vertical columns mark inelastic light scattering
lines. The rest of the spectrum is composed of the luminescence
bands.}
\end{figure}
\par
We focus here on the two cyclotron spin-flip modes SF1 and SF2
which are blue-shifted from $\hbar \omega_c$ by 1.13\,meV and
4.3\,meV respectively. We have compared these experimental values
to those calculated theoretically within the first-order
Hartree-Fock approximation taking into account the actual width of
the quantum well (see Fig.\,1c). The discrepancy is of the order
of the negative second order corrections such as for AP plasmon.
The Coulomb energy of SF2 is close to the estimated full exchange
energy of electrons on LL$_1$. The latter is represented by the
energy limit of shortwave SW at $\nu=3$. This asymptotical value
is about three fourth of analogous quantity at fully spin
polarized state $\nu=1$. We also find a marked dependence on
magnetic field in which lines SF1 and SF2 are observed only in the
narrow interval $\Delta B \simeq 0.3$\,T around $\nu=3$. Outside
this range of fields and filling factors the lines disappear from
the spectrum. From this fact we conclude that these excitations
are inherent to filling factor $\nu=3$.
\par
To summarize, by means of inelastic light scattering we have
observed and identified four inter-Landau level collective
excitations and intra-LL spin-wave at $\nu=3$. Among these
excitations there are two cyclotron spin-flip modes, which interact
repulsively in the long-wave limit. As a result, the upper of them
(SF2) acquires a huge exchange contribution to the energy,
comparable with the theoretically estimated exchange energy of
electrons on the first Landau level. The experimentally measured
energies of all excitations are in a good agreement with the
Hartree-Fock calculations taking into account the finite thickness
of the 2D-electron system.

\begin{acknowledgments}

The authors acknowledge support from the U.S. Civilian Research and
Development Foundation and the Russian Foundation for Basic
Research. T.D.R. and A. P. are supported by the National Science
Foundation under Grant No. DMR-0352738 and DMR-0803445, and by the
Nanoscale Science and Engineering Initiative of the National Science
Foundation under Award Number CHE-0117752 and CHE-0641523. T.D.R.
and A.P. were also supported by the Department of Energy under Grant
No. DE-AIO2-04ER46133, and by a research grant from the W. M. Keck
Foundation.
\end{acknowledgments}

\end{document}